# Electric field effects on the band gap and edge states of monolayer 1T'-WTe$_2$

Yulia Maximenko[1], Yueqing Chang[1], Guannan Chen[1], Mark R. Hirsbrunner[1], Waclaw Swiech[2], Taylor L. Hughes[1], Lucas K. Wagner[1] and Vidya Madhavan[1,2]

[1] Department of Physics, University of Illinois Urbana-Champaign, Urbana, Illinois 61801, USA
[2] Materials Research Laboratory, University of Illinois Urbana-Champaign, Urbana, Illinois 61801, USA

Monolayer 1T'-WTe$_2$ is a quantum spin Hall insulator with a gapped bulk and gapless helical edge states persisting to temperatures around 100 K. Recent studies have revealed a topological-to-trivial phase transition as well the emergence of an unconventional, potentially topological superconducting state upon tuning the carrier concentration with gating. However, despite extensive studies, the effects of gating on the band structure and the helical edge states have not yet been established. In this work we present a combined low-temperature STM and first principles study of back-gated monolayer 1T'-WTe$_2$ films grown on graphene. Consistent with a quantum spin Hall system, the films show well-defined bulk gaps and clear edge states that span the gap. By directly measuring the density of states with STM spectroscopy, we show that the bulk band gap magnitude shows substantial changes with applied gate voltage, which is contrary to the naïve expectation that a gate would rigidly shift the bands relative to the Fermi level. To explain our data, we carry out density functional theory and model Hamiltonian calculations which show that a gate electric field causes doping and inversion symmetry breaking which polarizes and spin-splits the bulk bands. Interestingly, the calculated spin splitting from the effective Rashba-like spin-orbit coupling can be in the tens of meV for the electric fields in the experiment, which may be useful for spintronics applications. Our work reveals the strong effect of electric fields on the bulk band structure of monolayer 1T'-WTe$_2$, which will play a critical role in our understanding of gate-induced phenomena in this system.

2D transition metal dichalcogenides present an exciting platform for realizing emergent phases having non-trivial topology and strong correlations, and are prime candidates for hosting topological edge states, topological superconductivity, and fractional excitations[1]. After more than a decade of intense search, monolayer $WTe_2$ (ML-$WTe_2$) (Fig. 1a) has recently emerged as a 2D time-reversal invariant topological insulator, i.e., a quantum spin Hall (QSH) insulator, and is now an important material for realizing and probing the associated 1D helical edge states[2,3,4,5]. While bulk $WTe_2$ is a semimetal, band structure calculations of a monolayer indicate that it is a narrow-gap semiconductor exhibiting inverted bands that give rise to the QSH phase (Fig. 1b)[6,7]. So far, transport studies have confirmed quantized two-terminal edge conductance up to 100 K, and microwave impedance microscopy measurements have been used to observe edge states[8,9].

An important control knob that is used to tune the properties of 2D materials is gating, i.e., the application of an electric field though a dielectric layer. In the simplest case, gating is expected to add or remove carriers from the film by shifting the bands with respect to the Fermi energy (rigid band shift) as is observed in graphene[10]. However, gating could have non-trivial effects on the band structure due to the presence of the out-of-plane electric field, or changes in effective screening. For example, out-of-plane electric fields break inversion symmetry and can lead to a spin splitting of the bands, like the Rashba effect. In $WTe_2$, electrostatic gating has been used to tune the circular photogalvanic effect[11] and induce an exotic superconducting phase[12,13]. Despite the far-ranging interest in this system and extensive transport studies, the exact gap size of ML-$WTe_2$, the dispersion of the edge states within the gap, as well as the effect of gating on the band structure are not established. Indeed, the experimental literature reports a puzzlingly large variation in the observed gap sizes, which has yet to be explained[2,3,5,7,8,11,12,13,14] (see Supplementary Fig. 3). Furthermore, this information has direct relevance to the role of topology and edge states in the insulating and superconducting phases. Hence, comprehensive gating-dependent spectroscopic data is urgently needed to understand the key elements that control the low-energy physics of the system, and scanning tunneling microscopy (STM) and spectroscopy (STS) are ideal tools to probe gate-induced effects on the band structure and edge states.

Unfortunately, barring a few exceptions, it has remained technically challenging to combine STM with gating. As an example, only back gates are compatible with an STM probe, and this makes it difficult to study thick films which screen out the gate-induced electric field. For our purposes, this means that studies of mono- or few-layer films are ideal since screening is minimized. Another issue lies with monolayer samples obtained by exfoliation which are usually micron sized. Making electrical contact to these samples, while keeping the surfaces clean for STM studies, has proven notoriously difficult. In this work we were able to overcome these challenges and carry out nanoscale spectroscopic measurements of the effects of gating on the quantum spin Hall (QSH) insulator monolayer 1T'-WTe$_2$. We achieve this by using molecular beam epitaxy (MBE) to grow monolayer films of WTe$_2$ on graphene substrates. We prepared ML-WTe$_2$ films on two different substrates: graphene-terminated SiC and graphene transferred onto SiO$_2$/doped-Si(111), which was used for the gating studies (see "Methods" section for details on film growth). Our method allows us to avoid exposing the sample to air and to avoid using photolithography to make contacts, thereby minimizing surface contamination.

Figure 1c shows a large scale STM image of ML-WTe$_2$ grown on graphene on SiC. Multiple monolayer islands of approximately 20-30 nm size, which correspond to the typical island size of MBE-grown WTe$_2$, can be seen in this field of view. Monolayer WTe$_2$ has a distorted orthorhombic 1T' structure which preserves inversion symmetry (Fig. 1a). Zooming into a representative island, we can clearly resolve the characteristic stripes corresponding to protruding Te atoms of the top surface (for example, Figs. 1d, e). The observed lattice corresponds well to 1T'-WTe$_2$ (Fig. 1a). A representative bulk spectrum averaged over an area of 5 by 5 nm$^2$ on one island (Fig. 1f) exhibits a band gap of approximately 40 meV.

As a benchmark for the topological electronic structure, we identify the bulk and edge states of an island (shown in Fig. 2a) with no applied gate voltage. Upon tracking the spectra from the center of the island to the edge, we observe a bulk insulating gap and an emergence of a finite density of states in the nominal insulating gap as the tip approaches the edge. The in-gap density of states arises from the helical edge modes expected for a QSH insulator. The edge modes can be seen more clearly in dI/dV maps shown in Fig. 2d, thus confirming the existence of helical QSH states which are highly localized to the edge near the Fermi energy. We find that the spatial extent of the edge modes increases as the energy of the modes approaches the bulk band edges. This trend can be seen on both sides of the Fermi energy as plotted in Fig. 2c. This is an expected feature since topological edge modes must eventually merge into the bulk bands at a point beyond which their localization length diverges[15,16]. In our data we find that the edge states are maximally localized at the Fermi level and merge into the bulk at the conduction band minima (CBM) and valence band maxima (VBM) values.

To characterize the ML-WTe$_2$ films grown on graphene on SiC further, we performed STM spectroscopy on multiple islands in this film. Surprisingly, we find that the bulk gap varies considerably from island to island (Figs. 3b-d). As measured by the difference between the CBM and VBM ($E_g=E_{cbm}-E_{vbm}$, see Supplementary Information), the bulk gap can vary between 0 meV and 60 meV for different micron-sized locations in a given film. This variation is not easy to immediately explain, especially since the topographies of areas exhibiting different gaps show no differences in structure or defect concentration. A recent study suggests that large strains can lead to gap variations in this system[17]. However, the samples in that study were deliberately grown to induce strain which is not the case for our samples. Another possible explanation is that the gap variations arise from trapped ions in the SiC substrate creating local electric fields. Such trapped ions have been previously observed in other films grown on SiC[18] and could result from the high energy RHEED electrons used to monitor film growth. In a nutshell, while our measurements substantiate earlier conflicting experimental determinations of the band gap magnitude, the question of the cause of the variations in gap magnitude remains open. Fortunately, as described below, using our gated-STM capability we can systematically investigate the origin of these gap variations by directly exploring the response of ML-WTe$_2$ to electric fields.

Our gating setup is shown in Fig. 4a where ML-WTe$_2$ was grown via MBE on chemical-vapor-deposited (CVD) graphene transferred onto 300 nm SiO$_2$/Si, with the Si substrate serving as the back gate electrode (see Supplementary Fig. 1 for details). Figures 4c and d show a topographical STM image of a representative WTe$_2$ island on graphene on SiO$_2$ including the clearly resolved atomic lattice. With this setup, we can apply positive and negative back gate voltages up to a maximum of 80 V. Naively, we expect that the applied electric field would simply modify the carrier concentration resulting in a rigid shift of the Fermi level relative to the bulk bands. The signature of this effect would be a rigid shift of the measured dI/dV spectrum relative to the Fermi energy, i.e., relative to zero STM bias. As shown in Fig. 4b, we do find that gating smoothly moves the Fermi energy towards the conduction/valence band consistent with electron/hole doping for positive/negative voltages respectively. However, we also find that gating has another, more dominant effect. As shown in Figs. 4b, e, and f, with increasing positive gate voltage the band gap decreases, while for negative gate voltages the band gap increases. Hence, we find an inherent sensitivity of the WTe$_2$ insulating gap to (even modest) electric fields. At face value it appears that positive gate voltages increase the band gap while negative voltages decrease the gap. However, we will show below that the discrepancy between positive and negative voltages is due to the fact that the STM signal is dominated by the density of states from the top layer of the sample. In reality both signs of the gate voltage decrease the band gap.

To gain an understanding of these gating-induced band structure changes, we carry out density functional theory (DFT) as well as model Hamiltonian calculations. We use DFT with the Heyd-Scuseria-Ernzerhof exchange-correlation functional (HSE06)[19] including spin-orbit interaction to compute the electronic structure of monolayer WTe$_2$. As noted in the methods section, we fit a tight-binding model to the DFT band structure using Wannier interpolation for our analysis. At zero gate electric field (see second panel of Fig. 5a), the conduction band minima are doubly degenerate at $k_\pm \simeq \pm \left(0.15\frac{2\pi}{a}, 0\right)$. At $k_+$, the spin ↑ electron (i.e., the spin pointing in the z-direction) is localized on the top Te atom due to strong spin orbit coupling, while the spin ↓ electron is localized on the bottom Te atom. Consistent with time-reversal symmetry, this is reversed in the $k_-$ direction, as noted on the diagram.

Next we model the applied electric field by adding a term ($H'$) to the tight-binding model, $H = H_0 + H' = H_0 + \sum_i eE \cdot z_i$, where $i$ indicates an orbital center. Fig. 5a shows the 2D bulk band structures for different applied gate electric fields, and we see that the electric field breaks the spatial inversion symmetry and spin-splits the doubly degenerate CBM at $k_\pm \simeq \pm \left(0.15\frac{2\pi}{a}, 0\right)$. It is clear that this splitting always *reduces* the (indirect) band gap, no matter the sign of the gate voltage. The spin-splitting is consistent with preserved time-reversal, but broken inversion symmetry. Explicitly we find that for negative gate voltage the conduction band minimum in the $k_+$ direction has spin ↓, while minimum in the $k_-$ direction has spin ↑. For the opposite electric field, this situation is reversed. In Fig. 5a, the bands are colored according to the projection of the electronic states onto the top Te atoms, which we expect to be the most accessible to the STM probe. For a negative gate voltage, the states in the conduction band near $k_\pm$ that are pushed downward by the spin splitting are primarily located on the top Te atoms while the states pushed upward are primarily on the bottom Te atoms. For positive gate voltage, the situation is reversed. Since we expect the STM to have a preferential coupling to the top Te atoms this scenario would predict that the STM-observed gap will decrease for negative gate and increase for positive gate, in clear alignment with the experimental results.

To directly compare with experiment, in Fig. 5b we plot the gap of the states projected on the top Te atoms as a function of electric field. The behavior of the measured gap with electric field, as shown in Fig. 4f, agrees well with the calculations shown in Fig. 5b. This also reconfirms the fact that the experimental observation of an increasing gap for positive gate voltages is due to the STM predominantly coupling to the top surface. For the 300 nm $SiO_2$ gate oxide, the 80 V gate voltage results in an electric field on the order of 100 mV/Å. A more accurate determination of the electric field strength is difficult because the thickness of the $SiO_2$ can show variations and graphene can induce some degree of screening. Nevertheless, using the value of 100 mV/Å as an estimate, we find that the change in the bulk gap (as measured from the top surface) to be 0.377(4) meV/mV, which is in excellent agreement with the experimental value of 0.39(2) meV/mV.

The calculated $dI/dV$ spectra at different electric field strengths is shown in Fig. 5c. The spectra are obtained by calculating the projected density of states (PDOS) of the Wannier functions of only the top Te atoms, i.e., $dI/dV|_\epsilon \propto \sum_i |\langle \phi_i | \psi \rangle|^2$, where $i$ runs through all the Wannier orbitals of the top Te atoms. The general trend of the experiment is reproduced by the DFT-derived tight-binding model, with the exception of a hump inside the gap at large positive fields, due to the fact that in the model calculations, the lower conduction band is slightly delocalized across both the top and the bottom Te atoms. This small discrepancy may be due to an underestimation of the localization in the DFT-derived model.

Finally, as an additional check, we confirmed this mechanism for the decrease in gap magnitude with gate electric field by considering a simple 6-band tight-binding model constructed from a crystal symmetry analysis informed from the low-energy band structure determined from DFT. Indeed, we find that the spin-splitting of the conduction band by the electric field is responsible for the observed behavior of the gap. We discuss this construction and explicit results in the Supplementary Information.

In summary, we have overcome numerous material challenges to achieve STM-accessible gating capabilities in $WTe_2$ monolayers. These devices offer great future potential for electric-driven effects in a topological insulator. Our work highlights the sensitivity of the band structure of monolayer 1T'-$WTe_2$ to electric fields, which cause substantial changes in the magnitude of the bulk band gap. Electronic structure calculations show that the effects due to electric fields can be primarily attributed to broken inversion symmetry and strong spin-orbit coupling akin to the Rashba effect rather than electron-electron interaction effects. Our systematic study, enabled by our newly developed gated-STM devices, resolve the puzzlingly large variations in the gap magnitudes that have been previously observed in monolayer 1T'-WTe$_2$. Indeed, we expect that the gap variations observed in the literature in the monolayer thin films may be attributed to electric fields arising from variations in the local potential induced by the substrate. Finally, it is worth noting that our studies indicate that WTe$_2$ is of potential interest for spintronic applications, since the gate electric field can be used to generate a substantial spin-splitting (DFT calculations show tens of meV) due to a Rashba-like spin-orbit coupling.

**Figure 1**

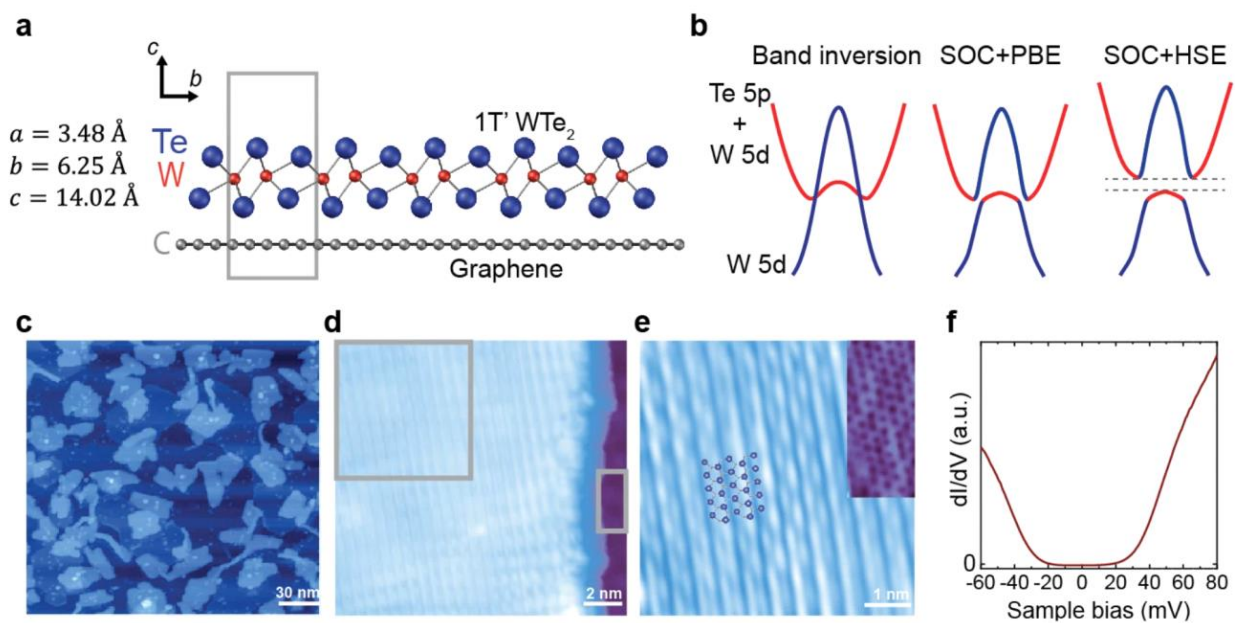

**Figure 1: Crystal structure and electronic structure of ML-WTe$_2$ films. a**, Crystal structure of monolayer 1T'-WTe$_2$. **b**, Schematic band structure of ML-WTe$_2$ (naive and DFT-calculated including spin-orbit coupling with PBE and HSE potentials respectively). **c**, A typical large-scale topography of MBE-grown monolayer WTe$_2$ on graphene on SiC. **d, e**, Zoomed-in topographies of monolayer islands showing characteristic Te stripes with atomic resolution. The graphene substrate can be identified to the right of the island. **f**, A typical STS spectrum for the monolayer bulk with a well-defined band gap of ~40 meV.

**Figure 2**

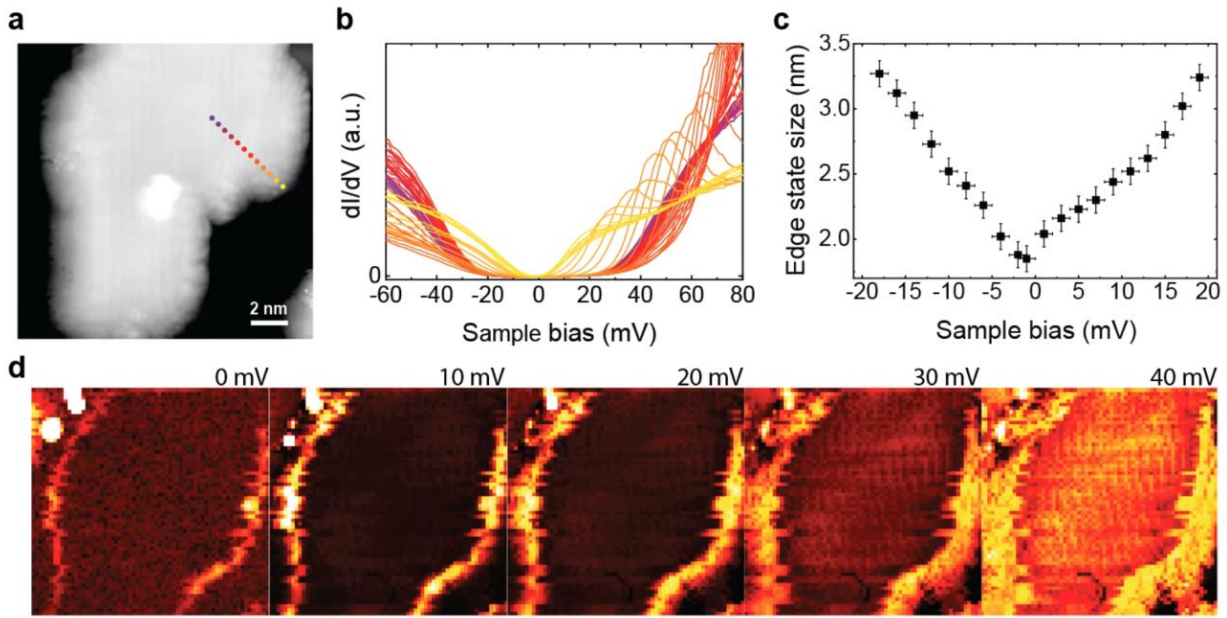

**Fig. 2 Edge states in ML-WTe$_2$ and their spatial characteristics. a**, Topography of one of the monolayer islands. The colored dots are the spatial locations where the corresponding spectra shown in **b** were obtained. **b**, dI/dV spectra the corresponding to the dot positions in **a**. **c**, Spatial extent of the edge state (a few representative energies are shown in **d)** for different STM biases. The extent is determined by choosing a dI/dV cut-off of 2% of the bulk spectrum value at 80 mV. The error bars were determined by the spatial resolution of the dI/dV map in **d**, and bias modulation used to obtain the dI/dV spectra. **d**, dI/dV spatially resolved local density of states maps of the island shown in **a** at the indicated energies. The maps show edge states continuous along the edge, at all energies inside the gap, with a spatial extent that increases with energy.

**Figure 3**

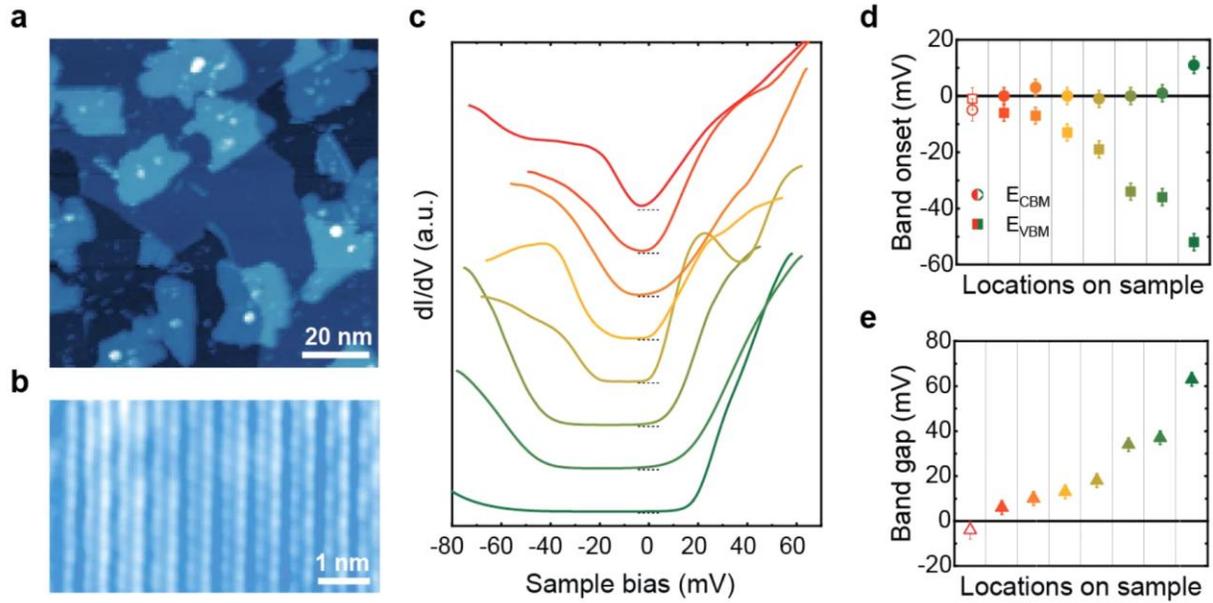

**Fig. 3 Band gap variations in ML-WTe$_2$ film grown on graphene on SiC. a**, Typical large scale STM topography of ML-WTe$_2$ films. **b**, Atomically resolved image of one of the monolayer islands. **c**, *dI/dV* spectra taken on the same sample arranged in the order of decreasing band gap. The spectra shown were taken from monolayer islands without any visible defects from different locations of the sample (see Supplementary Information). **d**, **e**, Conduction band minimum (CBM), valence band maximum (VBM) (**d**), and band gap (**e**) plotted for the spectra shown in **c**.

**Figure 4**

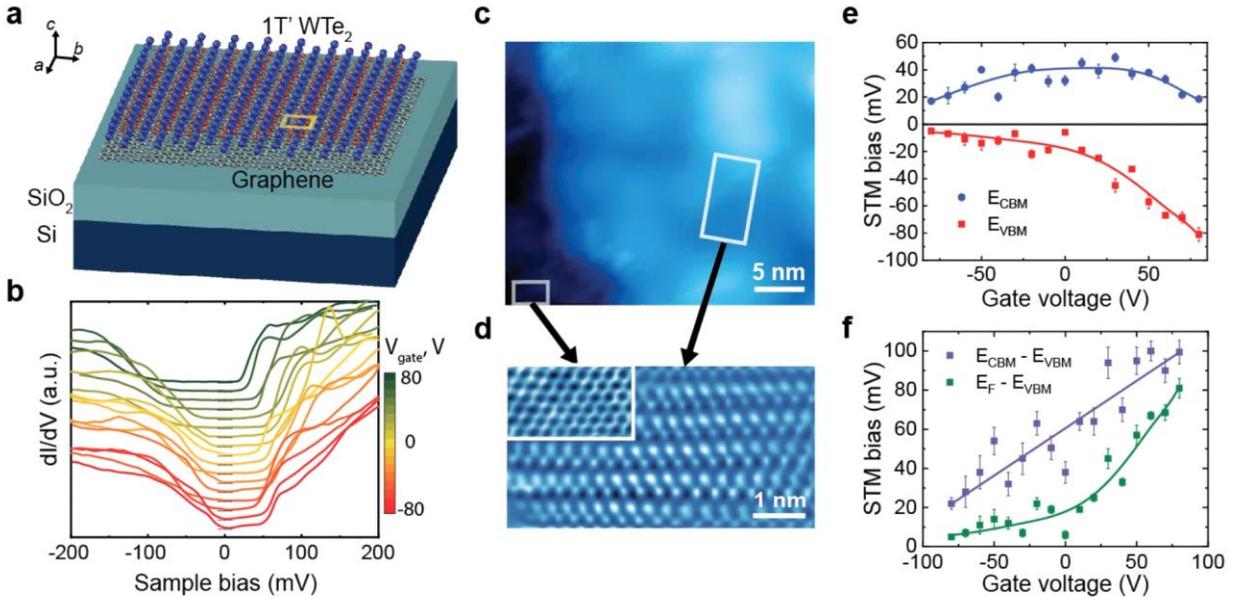

**Fig. 4 Back-gated STM measurements of ML-WTe$_2$. a**, Schematic of the back-gating setup. **b**, Evolution of the bulk d*I*/d*V* spectrum of a monolayer island in response to gate voltage. **c**, STM image of monolayer WTe$_2$ island on graphene on SiO$_2$. **d**, A zoom-in with atomic resolution clearly showing Te atoms as well as the graphene lattice (inset). **e**, Plots of the conduction band maximum and the valence band minimum for each spectrum at given gate voltage (blue and red dots respectively) showing the non-trivial evolution of the bulk electronic structure with gate voltage. The blue and red lines are guides to the eye. The error bars are determined by the energy intervals at which data were obtained and noise level of the *dI/dV* spectra. **f**, Same as **e** but with energies relative to the VBM. Back-gating changes the band gap (violet dots) as well as the position of the VBM with respect to E$_F$ (green dots). The violet and green lines are guides to the eye.

**Figure 5**

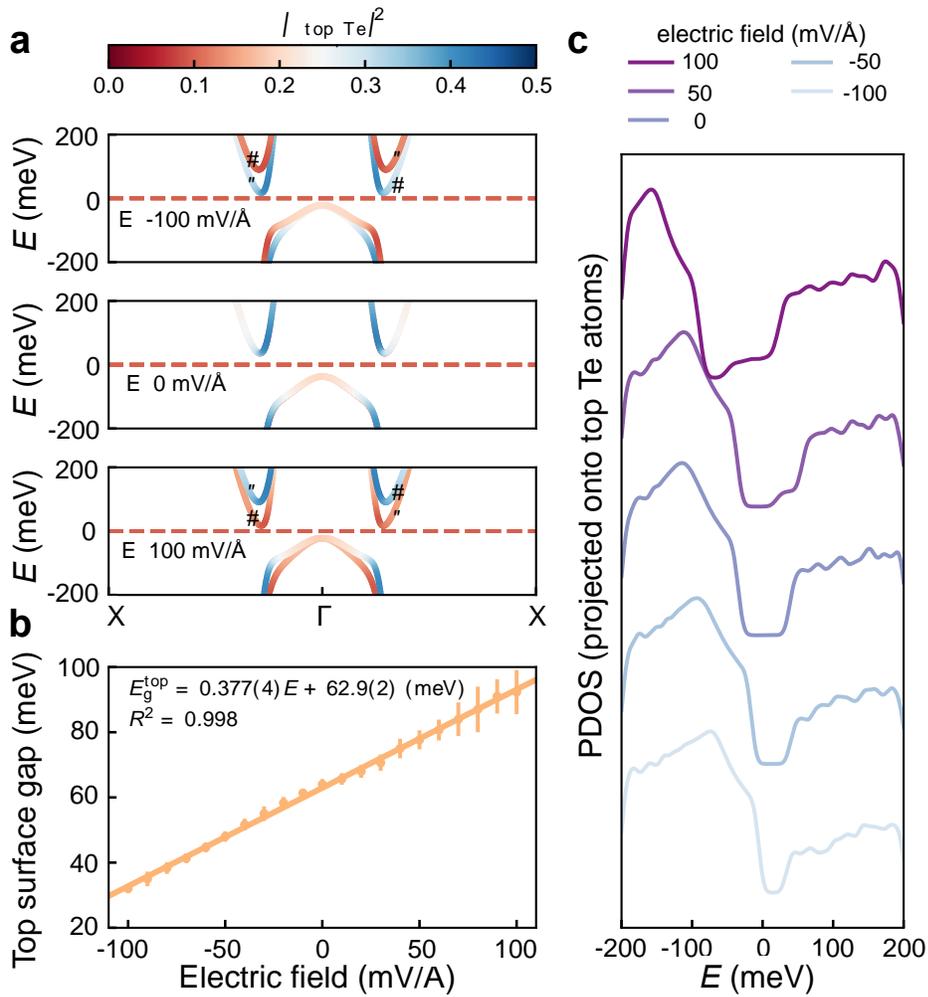

**Fig. 5 Theoretical calculations of the band structure and gating response of ML-WTe$_2$. a**, Band structure for -100, 0 and 100 mV/Å electric fields. We estimate that a gate voltage of 80V corresponds to approximately 100 mV/Å. The bands are colored with blue or red to represent electronic states more localized at the top Te atoms or the bottom Te atoms, respectively. The small black arrows indicate the spin polarization of the bands (near the band bottoms) due to the electric field from the applied gate voltage. **b**, The computed gap at the top surface plotted versus electric field. The slope and values are in close agreement with the STM data. **c**, Partial density of states (PDOS) with different gate voltages. The PDOS represents the density of states projected onto the top Te atoms.

**Methods**

Graphene-on-SiC substrates were fabricated directly on crystalline SiC in vacuum, providing a high-quality graphene surface for subsequent epitaxial growth. For gating studies, commercial graphene on $SiO_2$ substrates were used. These substrates are prepared by transferring monolayer CVD graphene onto a p-doped Si wafer coated with a 300 nm thermal $SiO_2$. Monolayer $WTe_2$ was grown using a custom MBE setup. Prior to growth, $Gr/SiO_2/Si$ substrates were annealed in UHV for 10 hours at 450°C. The film was grown by co-evaporation of elemental W (3N purity) and Te (6N purity) while the substrate was held at T = 280°C. Te was evaporated from a Kundsen cell with a rate of 0.01 Å/s. W was evaporated using an electron beam evaporator at a rate of 0.18 Å/hour. Monolayer film with 60% coverage was grown in five hours, thus having the rate of approximately one monolayer per 8.5 hours. Such slow growth and 200:1 Te to W flux ratio was necessary to grow high-quality monolayer films. After growth, the film was annealed in UHV at 300°C for one hour, cooled down to room temperature and capped with 15 nm of Te. The capped film was then transferred to the gating sample holder, which is a three-lead holder, one of which serves as the STM bias lead and the other two are connected to the heating W filament (see Supplementary Information). Once transferred to the STM system, the sample was ion-milled for 10 seconds using Ar+ beam at 400 V voltage and 4 µA beam current and then annealed at 200°C for one hour to remove the Te cap. The samples were immediately transferred into the STM stage kept at 4 K.

The density functional theory calculations are performed using Quantum ESPRESSO 6.5[20,21]. In order to compute the detailed band structure, we use Wannier interpolation. We derived a tight-binding model on a 56-spinor Wannier basis (W: *s* and *d*, Te: *s* and *p*) using Wannier90[22,23]. The tight-binding Hamiltonian is constructed and solved using the open-source code TBmodels[24].

# References

1. Manzeli, S., Ovchinnikov, D., Pasquier, D., Yazyev, O.V. & Kis, A. 2D transition metal dichalcogenides. *Nat. Rev. Mat.* **2**, 17033 (2017).
2. Fei, Z. et al. Edge conduction in monolayer $WTe_2$. *Nat. Phys.* **13**, 677–682 (2017).
3. Tang, S. et al. Quantum spin Hall state in monolayer 1T'-$WTe_2$. *Nat. Phys.* **13**, 683–687 (2017).
4. Jia, Z.-Y. et al. Direct visualization of a two-dimensional topological insulator in the single-layer 1T'-$WTe_2$. *Phys. Rev. B* **96**, 041108 (2017).
5. Song, Y.-H. et al. Observation of Coulomb gap in the quantum spin Hall candidate single-layer 1T'-$WTe_2$. *Nat. Commun.* **9**, 4071 (2018).
6. Qian, X., Liu, J., Fu, L. & Li, J. Quantum spin Hall effect in two-dimensional transition metal dichalcogenides. *Science* **346**, 1344–1347 (2014).
7. Zheng, F. et al. On the Quantum Spin Hall Gap of Monolayer 1T'-$WTe_2$. *Adv. Mater.* **28**, 4845–4851 (2016).
8. Wu, S. et al. Observation of the quantum spin Hall effect up to 100 kelvin in a monolayer crystal. *Science* **359**, 76–79 (2018).
9. Shi, Y. et al. Imaging quantum spin Hall edges in monolayer $WTe_2$. *Sci. Adv.* **5**, eaat8799 (2019).
10. Novoselov, K.S. et al. Electric field effect in atomically thin carbon films. *Science* **306**, 666–669 (2004).
11. Xu, S.-Y. et al. Electrically switchable Berry curvature dipole in the monolayer topological insulator $WTe_2$. *Nat. Phys.* **14**, 900–906 (2018).
12. Sajadi, E. et al. Gate-induced superconductivity in a monolayer topological insulator. *Science* ***362,*** 922–925 (2018).
13. Fatemi, V. et al. Electrically tunable low-density superconductivity in a monolayer topological insulator. *Science* **362**, 926–929 (2018).
14. Lüpke, F. et al. Proximity-induced superconducting gap in the quantum spin Hall edge state of monolayer $WTe_2$. *Nat. Phys.* **16**, 526–530 (2020).
15. Ok, S. et al. Custodial glide symmetry of quantum spin Hall edge modes in monolayer $WTe_2$. *Phys. Rev. B* **99**, 121105 (2019).
16. Lau, A., Ray, R., Varjas, D. & Akhmerov, A.R. Influence of lattice termination on the edge states of the quantum spin Hall insulator monolayer 1T'-$WTe_2$. *Phys. Rev. Mater.* **3**, 054206 (2019).

17. Zhao, C. et al. Strain tunable semimetal-topological-insulator transition in monolayer 1T′-WTe$_2$. *Phys. Rev. Lett.* **125**, 046801 (2020).
18. Krasheninnikov, A.V. & Nordlund, K. Ion and electron irradiation-induced effects in nanostructured materials. *J. App. Phys.* **107**, 071301 (2010).
19. Heyd, J., Scuseria, G.E. & Ernzerhof, M. Hybrid functionals based on a screened Coulomb potential. *J. Chem. Phys*. **118**, 8207 (2003).
20. Giannozzi, P. et al. QUANTUM ESPRESSO: a modular and open-source software project for quantum simulations of materials. *J. Phys. Condens. Matter* **21**, 395502 (2009).
21. Giannozzi, P. et al. Advanced capabilities for materials modelling with QUANTUM ESPRESSO. *J. Phys. Condens. Matter* **29**, 465901 (2017).
22. Pizzi, G. et al. Wannier90 as a community code: new features and applications. *J. Phys. Condens. Matter* **32**, 165902 (2020).
23. Mostofi, A.A et al. An updated version of wannier90: A tool for obtaining maximally-localised Wannier functions. *Comput. Phys. Commun*. **185**, 2309–2310 (2014).
24. Gresch, D. TBmodels documentation. https://tbmodels.greschd.ch/en/latest/ (2020).